\documentclass[conference]{IEEEtran}
\IEEEoverridecommandlockouts
\usepackage{cite}
\usepackage{amsmath,amssymb,amsfonts}
\usepackage{algorithmic}
\usepackage{graphicx}
\usepackage{textcomp}
\usepackage{xcolor}
\usepackage{float}
\usepackage{multirow}

\newcommand{\unit}[1]{\mbox{$\rm \,#1$}}
\newcommand{\Pst}{P_{st}}
\newcommand{\Plt}{P_{lt}}

\newcommand{\fc}{f_{c}}

\def\BibTeX{{\rm B\kern-.05em{\sc i\kern-.025em b}\kern-.08em
    T\kern-.1667em\lower.7ex\hbox{E}\kern-.125emX}}
\begin{document}
	
\IEEEoverridecommandlockouts
\IEEEpubid{\begin{minipage}{\textwidth}\ \\[10pt]
		\centering\footnotesize{\newline \newline \newline \newline \copyright 2022 IEEE. Personal use of this material is permitted. Permission from IEEE must be obtained for all other uses, in any current or future media, including reprinting/republishing this material for advertising or promotional purposes, creating new collective works, for resale or redistribution to servers or lists, or reuse of any copyrighted component of this work in other works. DOI:~10.1109/ICHQP53011.2022.9808778}
\end{minipage}}

\title{Decomposition Problem in Process of Selective Identification and Localization of Voltage Fluctuations Sources in Power Grids
\thanks{This research was funded in whole or in part by National Science Centre, Poland -- 2021/41/N/ST7/00397, the Foundation for Polish Science (FNP) -- Stipend START 45.2021. For the purpose of Open Access, the author has applied a CC--BY public copyright licence to any Author Accepted Manuscript ({AAM}) version arising from this submission.}
}

\author{\IEEEauthorblockN{Piotr~Kuwa{\l{}}ek}
\IEEEauthorblockA{\textit{Institute of Electrical Engineering and Electronics} \\
\textit{Poznan University of Technology}\\
Poznan, Poland \\
piotr.kuwalek@put.poznan.pl}

}

\maketitle

\begin{abstract}
Voltage fluctuations are common disturbances in power grids, therefore the effective and selective process of identification and localization of individual voltage fluctuations sources is necessary for the minimization of such disturbances. Selectivity in the process of identification and localization disturbing loads is possible by the use cascade of blocks: demodulation, decomposition and propagation assessment. The effectiveness of this approach is closely related to the used method of decomposition. The paper presents the problem of decomposition process for the selected method of selective identification and localization of voltage fluctuation sources, in which the algorithm of enhanced empirical wavelet transform ({EEWT}) is used as the decomposition method. The paper presents selected research results from the real power grid, for which the result of selected approach causes mistakes in the process of identification and localization of voltage fluctuations sources. The potential causes of such mistakes related to the decomposition process are discussed on the basis of obtained research results.
\end{abstract}


\section{Introduction}

Voltage fluctuations are one of disturbances in power quality. This phenomenon is defined as the continuous variation of r.m.s. or peak value of voltage~\cite{b1}. Currently, in many countries, the voltage fluctuations severity is evaluated by the flicker~\cite{b2} severity assessment with using the short-term~$\Pst$~\cite{b3} and long-term~$\Plt$~\cite{b4} flicker indicator. On the basis of the last benchmark report on power quality~\cite{b5}, voltage fluctuations are one of the most common phenomena disturbing power quality in power grids. Considering this fact, it is often necessary to take special operations~\cite{b6a,b6b,b6c} to minimize effects caused by voltage fluctuations. In real power grids, such operations are often preceded by the process of identification and then localization voltage fluctuations sources in power grids~\cite{b7a,b7b}. Identification is understood as the recognition type of disturbance source on the basis of extracted signal features, and localization is understood as the indication of supply point of disturbing load.

Methods of identification and localization voltage fluctuations sources are divided into single--point~\cite{b8a,b8b} and multi--point~\cite{b9a,b9b} methods in the literature. Single--point methods for identification and localization disturbing sources are the dominant part of the literature, due to the implementation simplicity. However, in most cases, such methods only allow for indication the side that is the main disturbing source. In the case when the main disturbing source is on the power grid side, a multi--step iterative procedure is necessary. Multiple-point methods are characterized by greater diagnostic possibilities, because such methods allow for selective identification and localization supply points of particular disturbing sources. In the case of these methods, the most important are: selectivity, i.e., the possibility of indication as many disturbing loads in power grids as possible on the basis of one simultaneous measurement series; and the possibility of considering every type of disturbing loads in power grids (including loads changing their operating state with a frequency greater than the power frequency~\cite{b10a,b10b}, e.g., power electronic devices~\cite{b11a,b11b}). One of solutions, which allow for the achievement of indicated goals, is the method of selective identification and localization of voltage fluctuations sources based on a cascade of blocks~\cite{b12}: demodulation~\cite{b13} -- decomposition~\cite{b14} -- propagation assessment~\cite{b15}. The method is conceptually complex, but allows for the selective and automatic localization of many potential disturbing loads without additional expert knowledge, which is not possible for other methods of identification and localization of voltage fluctuations sources, which are available in the literature. In the case of indicated method, the selectivity of identification and localization of disturbing loads depends on the used decomposition method, which carries out the task of automatic segmentation of analyzed signal into component signals associated with individual disturbing sources. Hence, it is important that the used decomposition method allows for the automatic adaptation of segmentation process to the analyzed signal, which can be noisy and non--stationary. Such decomposition properties are referred to as empirical.

In the next paper sections, the decomposition problem is presented in relation to the selected method of selective identification and localization of voltage fluctuation sources~\cite{b12}, in which the enhanced empirical wavelet transform ({EEWT})~\cite{b16} is chosen as the decomposition method. The paper presents selected measurement results from the real power grid, for which the selected method causes errors in the identification and localization of voltage fluctuations sources, which is directly related to the used decomposition method in the considered method~\cite{b12}. Potential causes of errors are discussed on the basis of obtained research results.

\section{Selected algorithm of selective identification and localization of voltage fluctuations sources}

\begin{figure}[!]\centering
	\includegraphics[width=\columnwidth]{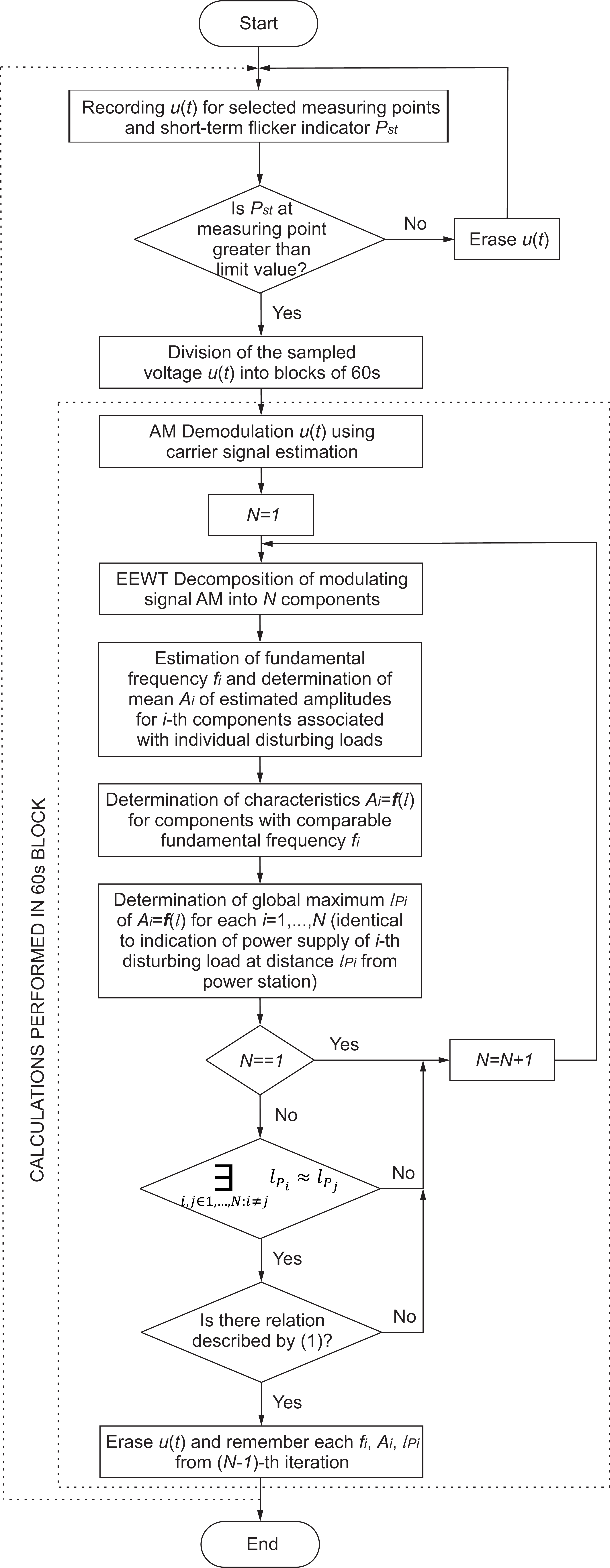}
	\caption{The considered algorithm for selective identification and localization of particular disturbing sources~\cite{b12}.}
	\label{alg}
\end{figure}

The selected algorithm of selective identification and localization of voltage fluctuations sources is presented in Fig.~\ref{alg}. The considered approach consists of three main blocks and is performed in an iterative process for 1--minute intervals. In each iteration, the estimation of real modulating signals is first realized by amplitude demodulation using carrier signal estimation~\cite{b13}. Secondly, estimated amplitude modulating signals are decomposed using enhanced empirical wavelet transform ({EEWT}) into $N$~components~\cite{b14}. For individual $i$--th components, the fundamental frequency $f_i$ and the average $A_i$ are determined from amplitudes of voltage changes~$\left( {{k}_{i}}^{\left( 1 \right) }, {{k}_{i}}^{\left( 2 \right) }, ...\right) $ (excluding outliers)~\cite{b12} for component signals after the regularization process~\cite{b12}. In the last step, the process of selective identification and localization of voltage fluctuations sources is performed by assessment of propagation of particular component signals, assuming that the propagation of modulating signal components proceeds in the same way as the propagation of resultant voltage fluctuations in power grids (resultant amplitude modulating signal)~\cite{b15}. Indication of supply points for particular disturbing loads is carried out by localization distance~$l_{{\text{P}}_i}$ for which the value of mean amplitude~$A_i$ of changes of $i$--th component signal (with a comparable fundamental frequency~$f_i$ for selected measurement points) achieves the global maximum value.

The iterative process starts with~$N = 1$ ($N$~is~the~number of decomposition signals that are associated with particular voltage fluctuations sources) and is repeated with an increase in~$N$ until any two $i$-th component signals result in the~indication of~the~same supply point. ($N$-1)--th iteration is important in the process of localization of supply points of particular disturbing loads. On the other hand, for the process of identification (recognition) of disturbing loads, the $N$--th iteration can be used (or subsequent iterations with an increase of $N$), provided that for the fundamental frequencies of any two component signals whose assessment of propagation results in the indication of the same supply point, relationship:
\begin{equation}\label{eq1}
	{{\exists }_{i,j\in 1,...,N:\left( i\ne j\wedge {{l}_{{{P}_{i}}}}\approx {{l}_{Pj}} \right)}}\left\{ {{f}_{i}}=\left| 2{{f}_{c}}-{{f}_{j}} \right|\vee {{f}_{i}}=n{{f}_{j}} \right\}
\end{equation}
is not satisfied, where $n \in \left\lbrace 1,2,3,4,5 \right\rbrace $, $\fc$~is the~power frequency, i.e., 50\unit{Hz} or 60\unit{Hz}.

\section{Research object}

\begin{figure}[htb]\centering
	\includegraphics[width=\columnwidth]{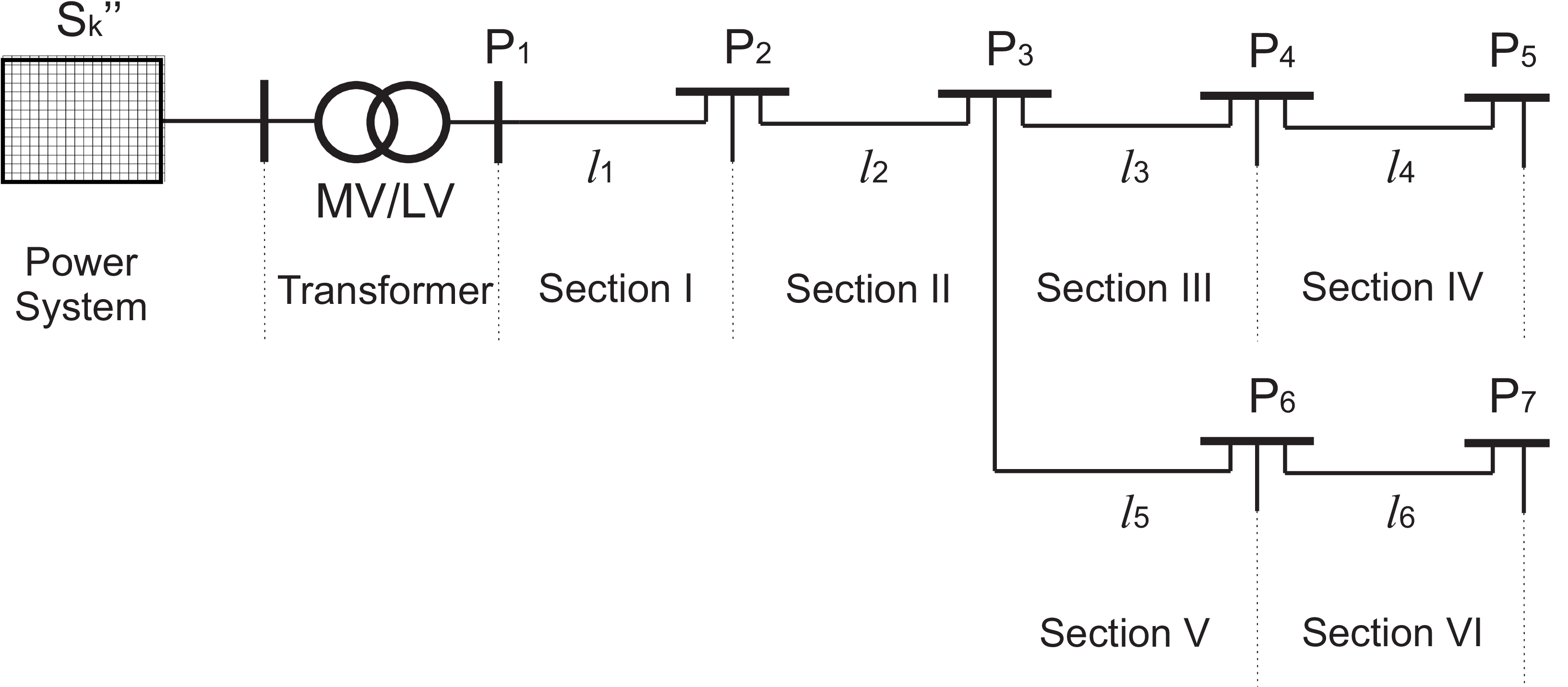}
	\caption{The considered low voltage power grid configuration with a branching radial topology.}
	\label{diagram_siec}
\end{figure}

The experimental studies were carried out on a prepared model of low voltage power grid with a branching radial topology, the diagram of which is shown in Fig.~\ref{diagram_siec}. The parameters of prepared power grid model are shown in Table~\ref{tab1}.

\begin{table}[!htb]
	\centering
	\caption{Parameters of considered low voltage power grid with branching radial topology}
	\label{tab1}
	\resizebox{\columnwidth}{!}{
		\begin{tabular}{c|c|c|c|c|c|c|c}
			Parameter    & SCP                                                & Type                                                          & \begin{tabular}[c]{@{}c@{}}${l}_{i}$\\ {[}m{]}\end{tabular} & \begin{tabular}[c]{@{}c@{}}S\\ {[}${\textnormal {mm}}^{2}${]}\end{tabular} & \begin{tabular}[c]{@{}c@{}}R\\ {[}${\textnormal m \Omega}${]}\end{tabular} & \begin{tabular}[c]{@{}c@{}}X\\ {[}${\textnormal m \Omega}${]}\end{tabular} & \begin{tabular}[c]{@{}c@{}}Z\\ {[}${\textnormal m \Omega}${]}\end{tabular} \\ \hline
			Power System & \begin{tabular}[c]{@{}c@{}}200 \\ MVA\end{tabular}  & -                                                             & -                                                    & -                                                     & 0.0                                                  & 0.9                                                  & 0.9                                                  \\
			Transformer  & \begin{tabular}[c]{@{}c@{}}630 \\ kVA\end{tabular} & -                                                             & -                                                    & -                                                     & 3.8                                                  & 10.8                                                 & 11.4                                                 \\
			Section I    & -                                                  & \begin{tabular}[c]{@{}c@{}}Overhead \\ line – Al\end{tabular}    & 300                                                  & 70                                                   & 150.0                                                 & 31.4                                                 & 153.3                                                 \\
			Section II   & -                                                  & \begin{tabular}[c]{@{}c@{}}Overhead \\ line – Al\end{tabular} & 250                                                  & 50                                                    & 150.0                                                 & 31.4                                                 & 153.3                                                \\
			Section III  & -                                                  & \begin{tabular}[c]{@{}c@{}}Overhead \\ line – Al\end{tabular} & 100                                                  & 35                                                    & 100.0                                               & 69.1                                                 & 121.6                                                \\
			Section IV   & -                                                  & \begin{tabular}[c]{@{}c@{}}Cable \\ line – Al\end{tabular}    & 50                                                   & 16                                                    & 50.0                                                 & 2.1                                                  & 50.0
			\\
			Section V   & -                                                  & \begin{tabular}[c]{@{}c@{}}Overhead \\ line – Al\end{tabular}    & 100                                                   & 35                                                    & 100.0                                                 & 69.1                                                  & 121.6
			\\
			Section VI   & -                                                  & \begin{tabular}[c]{@{}c@{}}Cable \\ line – Al\end{tabular}    & 50                                                   & 16                                                    & 50.0                                                 & 2.1                                                  & 50.0                                                
		\end{tabular}
	}
\end{table}

\begin{table}[!htb]
	\centering
	\caption{Parameters of disturbing loads for considered cases}
	\label{tab2}
	\resizebox{.9\columnwidth}{!}{
		\begin{tabular}{c|cccccc}
			\multirow{3}{*}{Case no.} & \multicolumn{6}{c}{Disturbing load}                                                                                                                             \\ \cline{2-7} 
			& \multicolumn{2}{c|}{VFS1 (2\unit{kW})}                            & \multicolumn{2}{c|}{VFS2 (3\unit{kW})}                            & \multicolumn{2}{c}{VFS3 (0.4\unit{kW})}      \\ \cline{2-7} 
			& \multicolumn{1}{c|}{$f_i$ {[}Hz{]}} & \multicolumn{1}{c|}{${\text{P}}_i$} & \multicolumn{1}{c|}{$f_i$ {[}Hz{]}} & \multicolumn{1}{c|}{${\text{P}}_i$} & \multicolumn{1}{c|}{$f_i$ {[}Hz{]}} & ${\text{P}}_i$ \\ \hline
			I                         & \multicolumn{1}{c|}{1.7}         & \multicolumn{1}{c|}{${\text{P}}_4$} & \multicolumn{1}{c|}{0.25}        & \multicolumn{1}{c|}{${\text{P}}_4$} & \multicolumn{1}{c|}{--}          & -- \\
			II                        & \multicolumn{1}{c|}{1.7}         & \multicolumn{1}{c|}{${\text{P}}_3$} & \multicolumn{1}{c|}{0.25}        & \multicolumn{1}{c|}{${\text{P}}_4$} & \multicolumn{1}{c|}{--}          & -- \\
			III                       & \multicolumn{1}{c|}{0.23}        & \multicolumn{1}{c|}{${\text{P}}_3$} & \multicolumn{1}{c|}{9.11}        & \multicolumn{1}{c|}{${\text{P}}_6$} & \multicolumn{1}{c|}{1.67}        & ${\text{P}}_4$ \\
			IV                        & \multicolumn{1}{c|}{108.8}       & \multicolumn{1}{c|}{${\text{P}}_6$} & \multicolumn{1}{c|}{91.2}        & \multicolumn{1}{c|}{${\text{P}}_3$} & \multicolumn{1}{c|}{8.8}         & ${\text{P}}_4$ \\
			V                        & \multicolumn{1}{c|}{0.7}         & \multicolumn{1}{c|}{${\text{P}}_6$} & \multicolumn{1}{c|}{0.1}         & \multicolumn{1}{c|}{${\text{P}}_3$} & \multicolumn{1}{c|}{2.5}         & ${\text{P}}_4$
		\end{tabular}
	}
\end{table}

\begin{figure}[htb]\centering
	\includegraphics[width=\columnwidth]{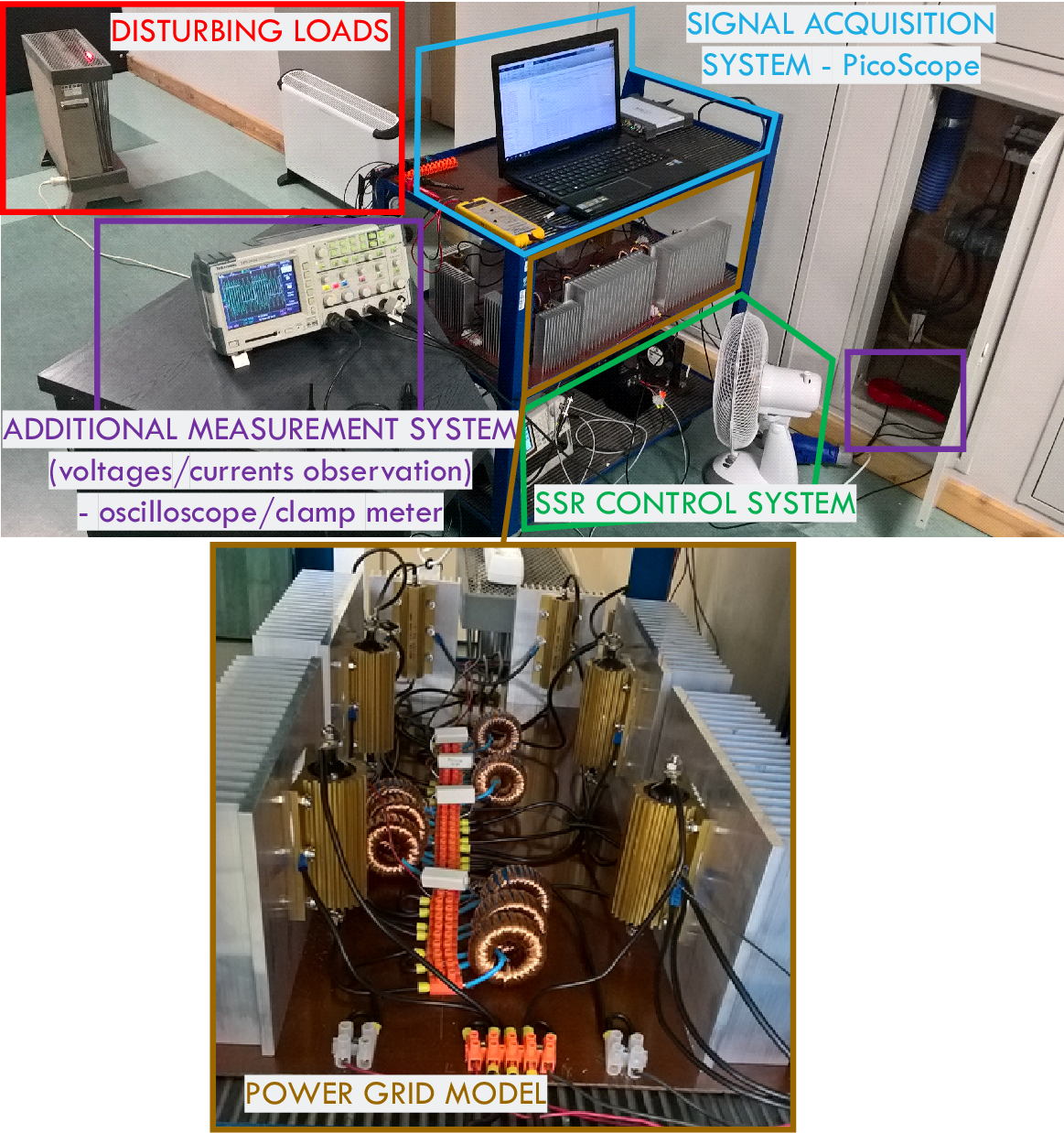}
	\caption{The laboratory setup for considered power gird model.}
	\label{lab_setup}
\end{figure}

Fig.~\ref{lab_setup} shows a laboratory setup for the considered model of power grid. In experimental studies, voltage fluctuations sources were: 2\unit{kW} convection--radiation heating system~({VFS1}), 3\unit{kW} ({VFS2}) and 0.4\unit{kW} ({VFS3}) convection heating systems, where each system was separately controlled by {SSR} systems with a (known) given frequency~$f_i$. Supply points ${\text{P}}_i$ of considered voltage fluctuations sources were known. Selected information on used voltage fluctuations sources used is included in Table~\ref{tab2}. Cases I--V were considered for which used the algorithm presented in Fig.~\ref{alg} (the sampling frequency in experimental studies was 12.5\unit{kHz}).

\section{Research results and discussion}

\begin{figure}[!htb]\centering
	\includegraphics[width=.9\columnwidth]{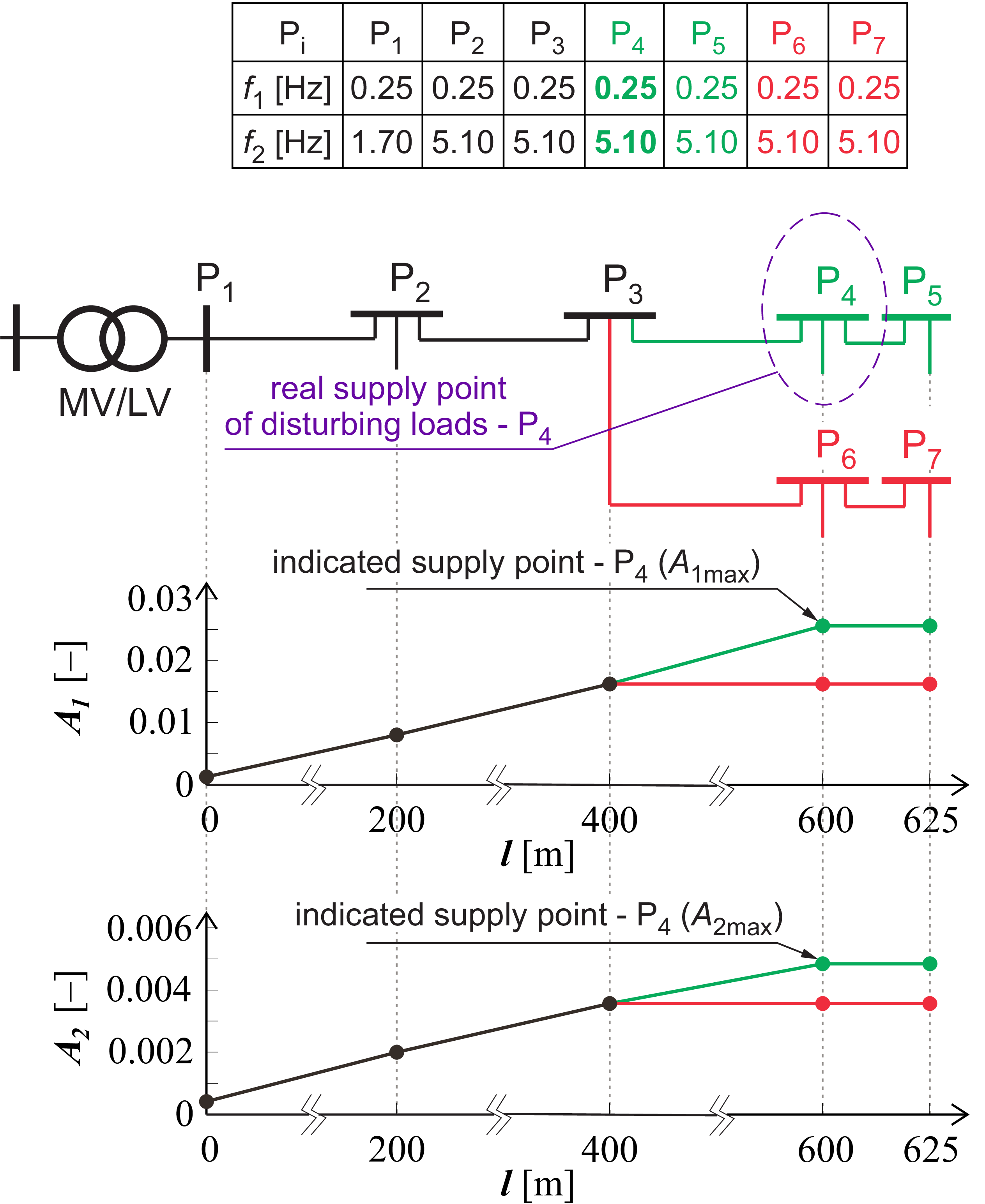}
	\caption{The graphical representation of considered algorithm for Case I.}
	\label{case_I}
\end{figure}

\begin{figure}[htb]\centering
	\includegraphics[width=.9\columnwidth]{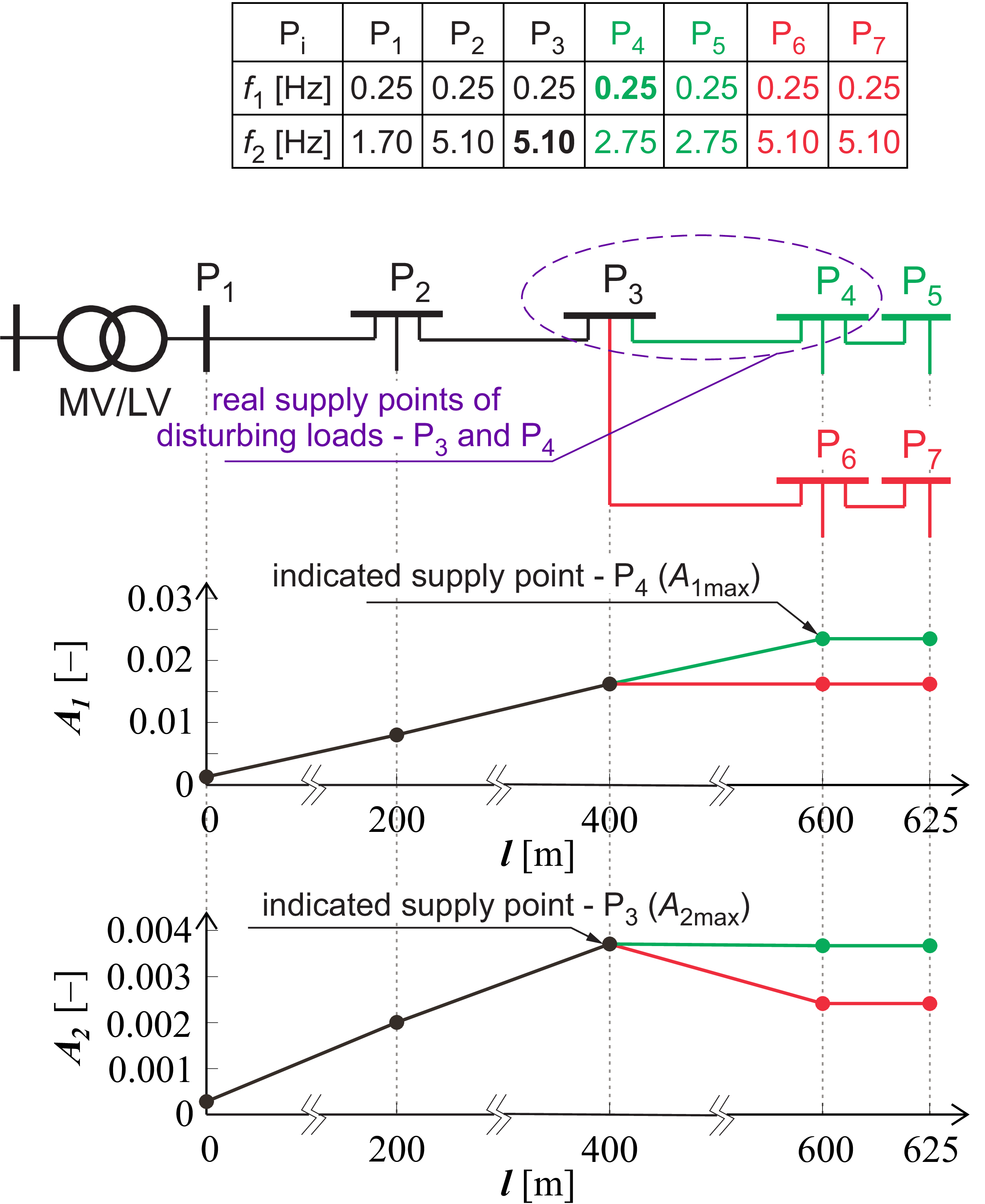}
	\caption{The graphical representation of considered algorithm for Case II.}
	\label{case_II}
\end{figure}

\begin{figure}[!htb]\centering
	\includegraphics[width=.9\columnwidth]{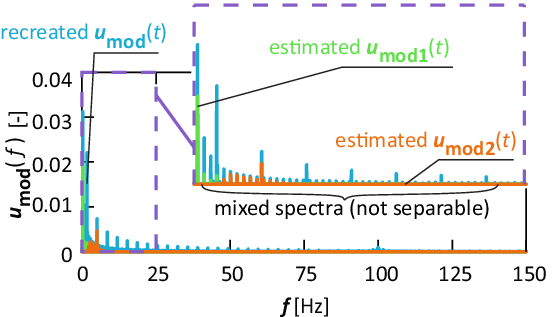}
	\caption{Spectra for the recreated modulating signal and for its individual component signals obtained in the process of decomposition by {EEWT} and regularization.}
	\label{widma}
\end{figure}

Fig.~\ref{case_I} and Fig.~\ref{case_II} show results of the algorithm shown in Fig.~\ref{alg} for Case~I and Case~II, respectively. On the basis of obtained research results, it can be seen that for considered cases, the correct assessment of number of disturbing sources is obtained and supply points of individual disturbing loads are correctly indicated. However, the frequency of operating state changes of one of voltage fluctuation sources is improperly estimated. The estimation error results from incorrect segmentation of spectra of analyzed signals during the construction of wavelet filter banks in the used decomposition method. As a result of the incorrect segmentation of spectrum, the cutoff frequency values of one of filters (constructed in such process) included: a significant range of non--negligibly small components of spectrum of component signal associated with the source VFS2; and a fundamental component of spectrum of component signal associated with the source VFS1. Remaining components of spectrum of component signal associated with the source VFS1 occurred in the pass--bandwidth of second constructed filter. The discussed situation is presented in Fig.~\ref{widma}.

\begin{figure}[htb]\centering
	\includegraphics[width=.9\columnwidth]{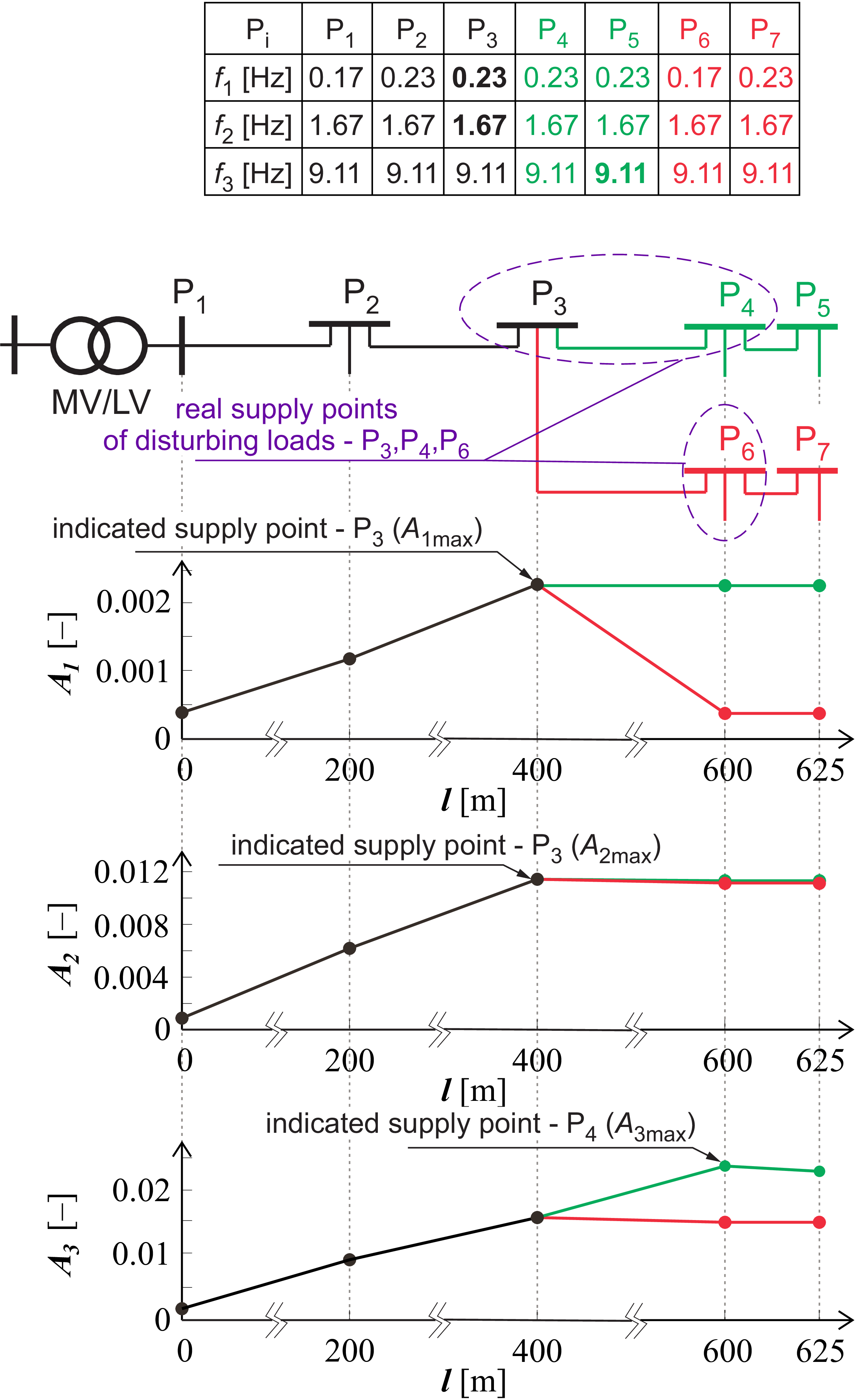}
	\caption{The graphical representation of considered algorithm for Case III.}
	\label{case_III}
\end{figure}

\begin{figure}[htb]\centering
	\includegraphics[width=.9\columnwidth]{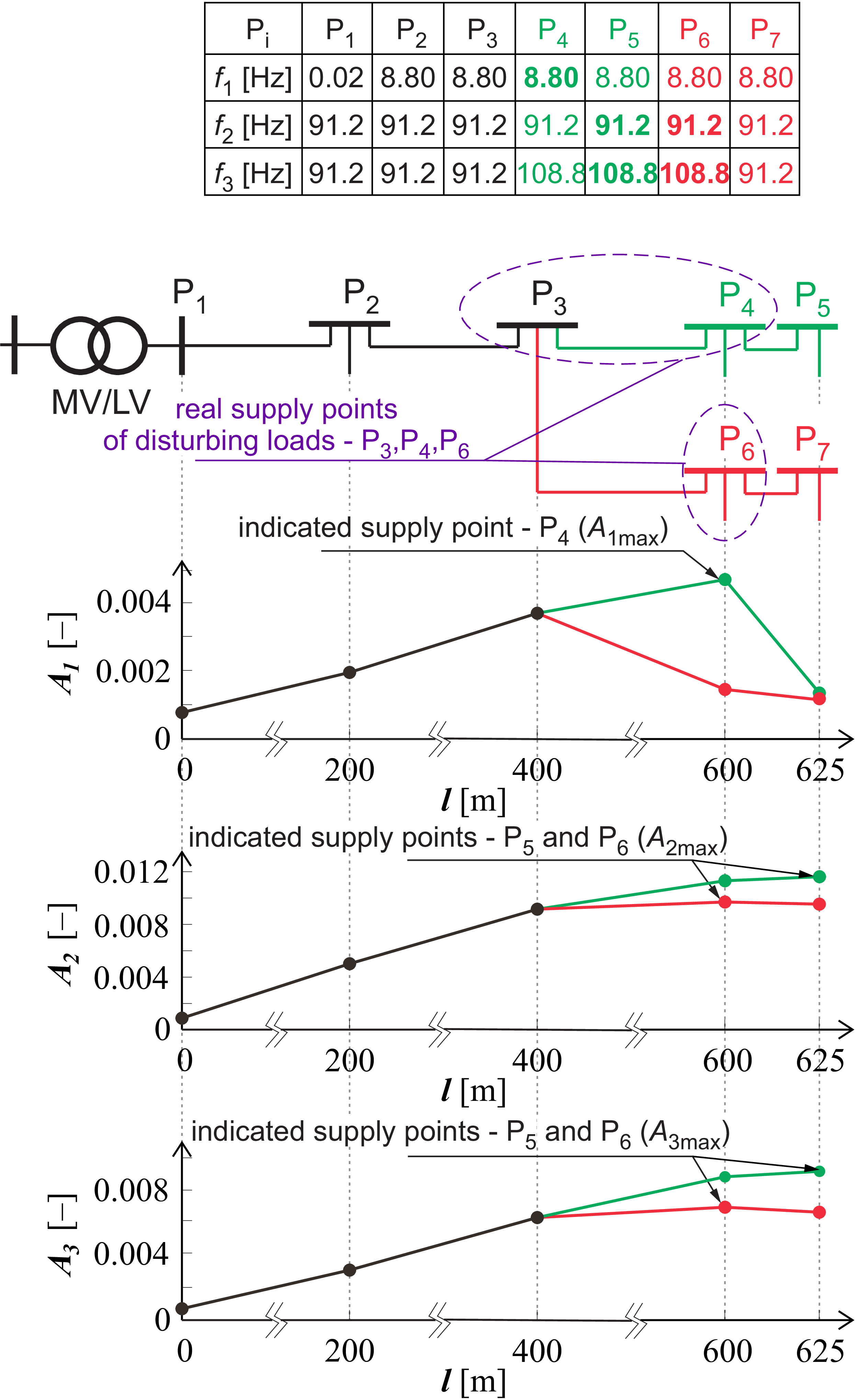}
	\caption{The graphical representation of considered algorithm for Case IV.}
	\label{case_IV}
\end{figure}

\begin{figure}[htb]\centering
	\includegraphics[width=.9\columnwidth]{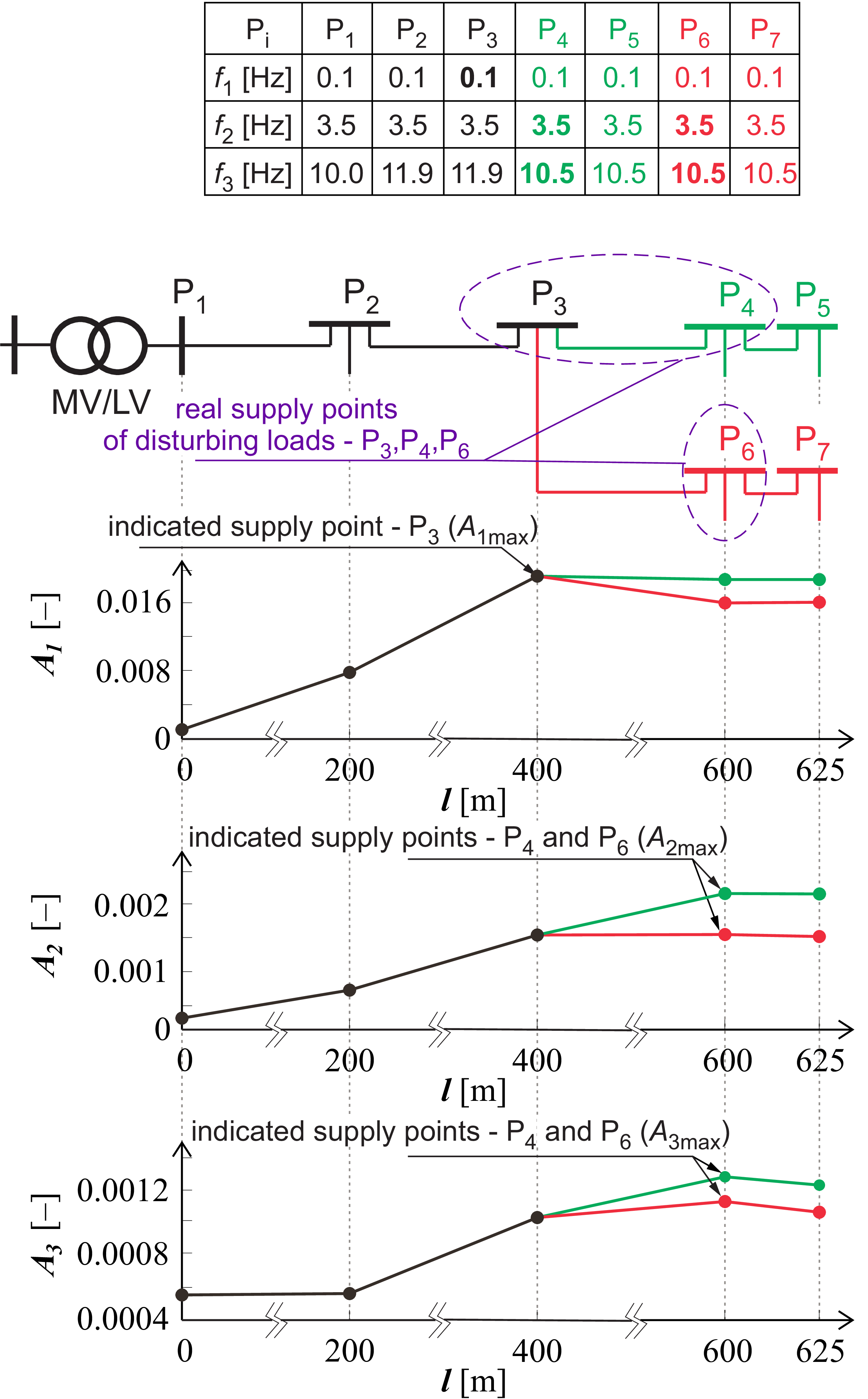}
	\caption{The graphical representation of considered algorithm for Case V.}
	\label{case_V}
\end{figure}

\begin{figure}[htb]\centering
	\includegraphics[width=.99\columnwidth]{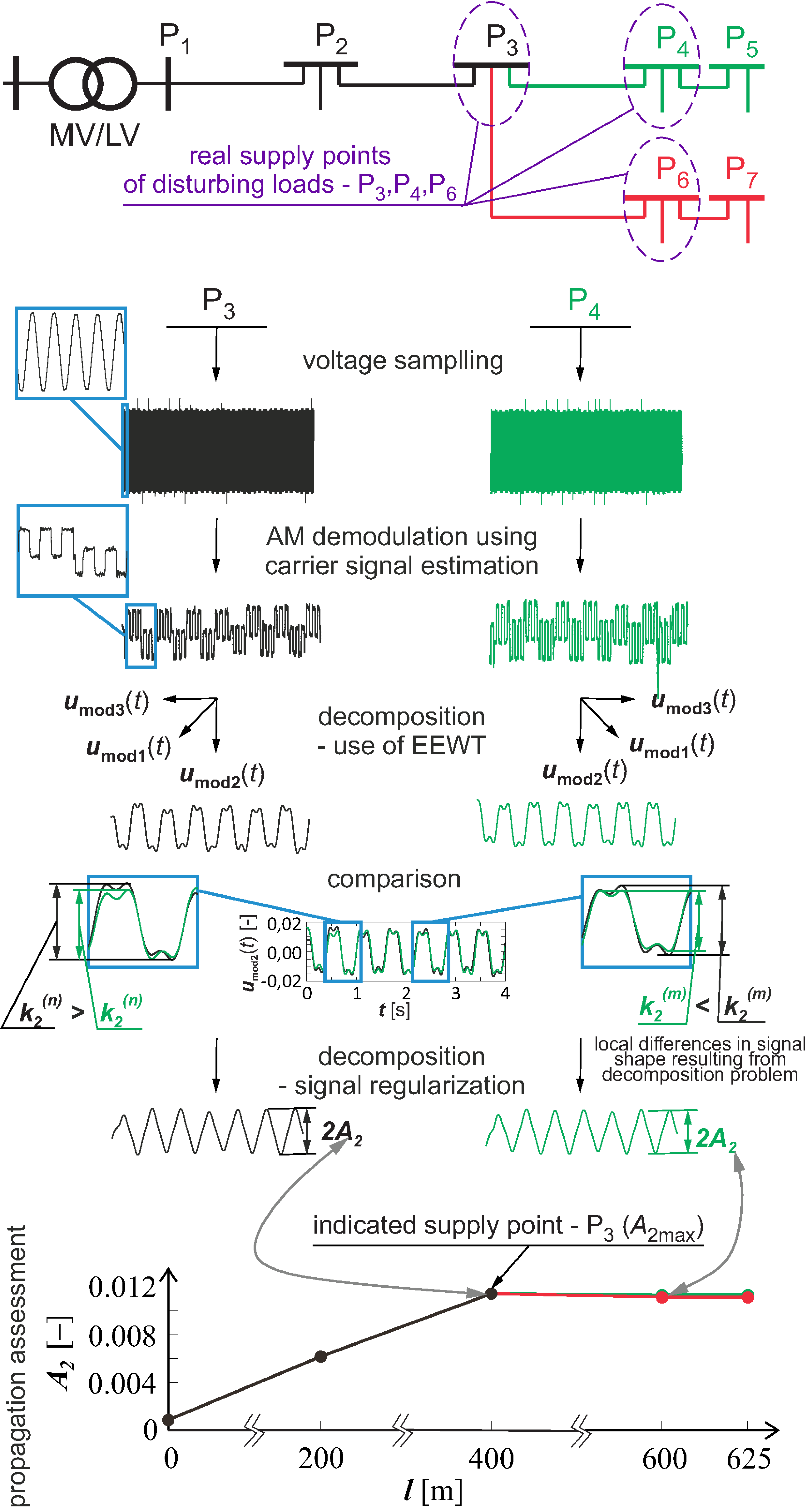}
	\caption{The graphical representation of individual stages of signal processing (for supply points ${\text{P}}_3$ and ${\text{P}}_4$) in the considered algorithm for case V.}
	\label{czasowe}
\end{figure}

Figs.~\ref{case_III}--\ref{case_V} show results of the algorithm shown in Fig.~\ref{alg} for Cases III--V, respectively. On the basis of obtained research results, it can be seen that for considered cases, the correct assessment of number of disturbing sources is obtained. However, depending on the case, the frequency of changes in the operating state of some voltage fluctuation sources is improperly estimated and the supply point of some voltage fluctuation sources is incorrectly indicated. Incorrect indications of supply points of individual disturbing loads, as in the case of identification mistakes, result from incorrect segmentation of spectra of analyzed signals during the construction of wavelet filter banks of used decomposition method. As a result of incorrect spectrum segmentation, the spectrum components of individual component signals of recreated modulating signal are contained in the pass--bandwidth of individual constructed filters (component signals are not separated into independent component signals). As a result, determined means of amplitudes of voltage changes are not directly associated with individual voltage fluctuations sources and, depending on the construction of wavelet filters, the estimated component signal has a different shape resulting from the fact that the signal is recreated by the geometric (not algebraic) sum of spectral components of individual modulating signal components. For a correctly implemented process of decomposition into independent component signals associated with individual disturbance sources, shapes of component signals with comparable fundamental frequency should be the same. Differences can occur in amplitudes, which depend on the properties of disturbance source and the properties of power supply circuit~\cite{b15}. The discussed situation is presented in Fig.~\ref{czasowe}.

It is worth noting that for considered disturbing sources, component signals have a quasi--rectangular shape with asymmetric duty cycle, for which the spectrum is complex. If component signals are:
\begin{itemize}
	\item[-] sinusoidal/quasi--sinusoidal signals; or
	\item[-] signals for which spectra have such property that the most of spectrum energy is concentrated near the fundamental frequency; or
	\item[-] signals for which spectra are approximately separable (or at least there is a significant distance between the fundamental components of spectra for individual component signals~\cite{b12});
\end{itemize}
then there are no problems of selective identification and localization of voltage fluctuations sources~\cite{b12}, which are discussed for considered Cases I--V.

\section{Conclusion}

The paper presents the preliminary results of research on decomposition problems in the process of selective identification and localization of voltage fluctuations sources in power grids, which can result in an incorrect estimation of frequency of changes in the operating state of individual disturbing sources, or an incorrect indication of the supply point of individual disturbing sources. Such situation can occur if component signals of recreated amplitude modulating signal have a complex spectrum that mixes with other spectra, causing the problem of correct separation. The presented decomposition problem related to the enhanced empirical wavelet transform is discussed on the basis of research results obtained for the prepared power grid model supplied directly from the real power grid. In practice, component signals of the amplitude modulating signal (associated with operations of voltage fluctuations sources) can have different undefined shapes due to the random nature of voltage fluctuation sources. Hence, it is difficult determination relationship that should occur between individual component signals for the correct and effective process of selective identification and localization of disturbing sources. It is possible that the presented problem of decomposition by {EEWT} could be eliminated by using other methods of empirical decomposition or by proposition of new decomposition method with basis functions associated with amplitude modulating signals most often occurring in real power grids.

\vspace{12pt}

\end{document}